\documentclass[10pt]{amsart}
\usepackage{lineno}
\usepackage{amssymb, amsfonts, amsthm, dsfont,bm}

\newtheorem{defi}{\textbf{Definition}}[section]
\newtheorem{thm}{\textbf{Theorem}}[section]

\newtheorem{lemma}{\textbf{Lemma}}[section]
\newtheorem{rem}{\textbf{Remark}}[section]

\newtheorem{cond}{\textbf{Condition}}

\def\nn{\mathbb{N}_{0}}
\def\rr{\mathbb{R}}

\def\P{\mathcal{P}}

\def\Per{\mathrm{Per}}

\def\dd{\textup{d}}

\title[Markov approximations of  Gibbs measures]
{Markov approximations of Gibbs measures for long-range potentials on 1D lattices}
\author[C. Maldonado \& R. Salgado-Garc\'{\i}a]{Cesar Maldonado$^{*}$ \& Ra\'{u}l Salgado-Garc\'{\i}a$^{\dagger}$}
\address{*Instituto de F\'{i}sica, Universidad Aut\'{o}noma de San Luis Potos\'{\i}, 
Av. Manuel Nava 6, Zona Universitaria, 78290, San Luis Potos\'{\i} S.L.P., M\'{e}xico.} 
\address{e-mail address: \textit{\texttt{coma@dec1.ifisica.uaslp.mx}}}
\address{$\dagger$ Facultad de Ciencias, Universidad Aut\'onoma del Estado de Morelos. 
Avenida Universidad 1001, Colonia Chamilpa, 62209, Cuernavaca Morelos, Mexico. }
\address{e-mail address: \textit{\texttt{raulsg@uaem.mx}}}

\thanks{The authors thank Edgardo Ugalde for introducing us to this 
interesting topic of research, for his suggestions, corrections and interesting discussions. 
C.M. was supported by a postdoctoral fellowship PROMEP, UASLP-CA-188.
}
\date{}

\begin{document}

\begin{abstract}
We study one-dimensional lattice systems with pair-wise interactions of infinite range. 
We show projective convergence of Markov measures to the unique equilibrium state. 
For this purpose we impose a slightly stronger condition than summability of variations 
on the regularity of the interaction. With our condition we are able 
to explicitly obtain stretched exponential bounds for the rate of mixing of the equilibrium state.
Finally we show convergence for the entropy of the Markov measures to that of the equilibrium state via the convergence of their topological pressure.
\end{abstract}
\maketitle


\section{Introduction}

Two concepts that are of fundamental importance in the rigorous study of statistical mechanics are the Gibbs measures and the equilibrium states. 
Many important results concerning the latter within the thermodynamical formalism, were first established in lattice systems and 
then generalized to  much more abstract settings such as  topological dynamical systems.
The concept of equilibrium state as a translationally invariant measure was  introduced 
within study of  the statistical mechanics of $d$-dimensional lattice systems
mainly motivated by the occurrence of the phenomenon known as the \emph{spontaneous 
symmetry breakdown}~\cite{dobrushin1968problem, LR}. In late 60's Bowen and  Ruelle~\cite{Bow2, Ruelle} independently introduced the \emph{variational principle} to characterize equilibrium states. On the other hand, for lattice systems, Gibbs measures associated to sufficient regular potentials are known to be equivalent to equilibrium states~\cite{LR, dobrushin1968gibbsian}. It is due to Bowen its characterization from the modern point of view~\cite{Bow}.

In this paper we are concerned with the problem of uniqueness of Gibbs measures in 
a one-dimensional lattice with long range pair interactions. On this class of systems several conditions for the existence 
and uniqueness of Gibbs measures has been established since its introduction. 
For instance, Ruelle gave sufficient conditions on the regularity of the interaction 
to prove the uniqueness of the equilibrium state. 
Then, Bowen extended such results to general dynamical systems defined by expansive homeomorphisms on compact metric spaces. 
He proved also that the unique equilibrium state is partially mixing. 
Years later, but in the same context, Walters~\cite{Walters} proved the Ruelle's convergence 
theorem for $g$-measures using the operator techniques proposed in~\cite{Ruelle}. 

Under the Bowen-Walters' condition, uniqueness of the equilibrium state has been 
proved following several approaches~\cite{Bow2,dobrushin1968problem,Ruelle,Walters}. The classical argument, which can be found in~\cite{dobrushin1968problem} for instance, goes as follows. 
First one defines an atomic measure on a periodic sequence, playing the role of a finite-volume measure in the canonical ensemble. Then one 
shows that the accumulation points as the period goes to infinity, are equilibrium states. 
In this way the existence part is proven. To prove uniqueness it is shown that equilibrium states 
are Gibbs measures under Bowen-Walters' condition. Then, by ergodic decomposition theorem it follows uniqueness since any two Gibbs measures are absolutely continuous with respect to each other in dimension one. Different conditions has been proposed to prove uniqueness since then~\cite{Sten}. 

More recently the authors of \cite{ChRU} proposed a technique of approximation of Gibbs
measures for H\"older-continuous potentials on sofic subshifts by means of Markov approximations on sofic subshifts. 
The strategy is based on an explicit construction of a sequence of periodic Gibbs measures for H\"older-continuous potentials on subshifts of finite type ``enveloping'' the sofic subshift. Then it is shown that the sequence of periodic measures has an unique accumulation point in the weak-$*$ topology  (via the Perron-Frobenius theorem), when both, the period goes to infinity and the subshift of finite type tends to the sofic subshift. Such an accumulation point is actually the unique Gibbs measure associated to the H\"older-continuous potential on the sofic subshift.

%
\noindent
With that technique they are able to obtain upper bounds for the rate of mixing of the limiting measure, as well as to prove convergence of the entropy.

Here we use the same technique of \cite{ChRU, ChU} in order to prove similar results but in the context of one-dimensional lattice systems with long-range potentials. Within this context it is possible to enhance the rate of mixing of the limiting measure.

The paper is organized as follows. Section 2 contains the necessary definitions, generalities and notational remarks. Section 3 establishes our main results: $i$) the Markov approximations converge to the unique equilibrium state in the projective sense, $ii$) an explicit upper bound for the rate of mixing of the limiting measure and $iii$) the convergence of the entropies of the Markov approximations to the entropy of the equilibrium state. Section 4 is devoted to develop explicitly the proofs of theorems and lemmas stated.

\section{Setting}

\subsection{Generalities}
We consider a one-dimensional infinite lattice. At each site of the lattice is placed a particle that interact with the other particles via pair interactions. In order to model such pair interactions between particles let us consider a finite set $A$, called alphabet, representing the ``state'' of the particle. As usual, the interaction of every pair of particles will depend on the specific states in which they are and the distance between them. Let $X=A^{\nn}$ be the space 
of all infinite sequences with symbols taken from $A$, in other words, the set of all possible configurations of our system. A word of length $k$ ($k\in \nn$) is the finite sequence 
$a_{1}^{k}:=a_{1}a_{2}\cdots a_{k}$ where $a_{i}\in A$ for all $i=1,\ldots,k$.
We endow $A$ with the discrete topology. 
The space $X$ is supplied with the product topology, and thus it is compact. We consider 
the Borel sigma-algebra generated by the cylinder sets defined by 
$[a_{0}^{k}] := \{ x\in X\ :\ x_{0}^{k}=a_{0}^{k} \}$. 

Next, the dynamics in $X$ is generated by the shift map $T$, that is, $(Tx)_{i} = x_{i+1}$ for any $x\in X$. 
Let $\mathcal{M}_{T}(X)$ be the space of all $T$ invariant probability measures on $X$. 
Consider the following distance on $\mathcal{M}_{T}(X)$.
\begin{equation*}
D(\mu, \nu) := \sum_{k=0}^{\infty} 2^{-{k+1}}\Big(\sum_{a\in A^{k}}\big\lvert \mu[a]-\nu[a] \big\rvert\Big),
\end{equation*}
it is well known that this distance is a metric for the weak-$*$ topology and
on the same topology $\mathcal{M}_{T}(X)$ is a compact and convex set.\\

\textbf{Notational remark:} From now on, we denote by $a = c^{\pm1}$ the inequalities $c^{-1}\leq a \leq c$. Analogously $a = \exp(\pm b)$ stands for $\exp(-b) \leq a\leq \exp(b)$. 

The following definition will be important for stating our results.

\begin{defi}[Projective convergence\footnote{This definition has been proposed by E.Ugalde and L.Trejo in \cite{liliana}.}]
Let $r,n\in\nn$. The sequence of measures $\{\nu_{r}\}$ is said to converge 
in the projective sense to $\nu$ if there exists a sequence $\{\varepsilon_{r}\}$, such that 
$\varepsilon_{r}\to 0$ as $r\to \infty$ for any cylinder $[a_{0}^{n-1}]$, for which the following 
inequalities hold
\[
\frac{\nu_{r}[a_{0}^{n-1}]}{\nu[a_{0}^{n-1}]} = e^{\pm n\varepsilon_{r}}.
\]
\end{defi}

\subsection{Pair-wise interactions}
We are interested in quantifying the ``energy'' of any configuration $x$. 
The local energy (centered at 0) of the configuration $x$ is the function 
$\phi:X\to\rr$ defined as follows,
\begin{equation}\label{poten}
\phi(x) := \sum_{k=0}^{\infty}\psi_{k}(x_{0},x_{k}),
\end{equation}
where $\psi_{k}:A\times A\to \rr$ is a distance-dependent pair-wise interaction.
\noindent
A classical example is an Ising-type model studied by Dyson, for which $A=\{ 1,-1\}$ and the pair-wise interaction 
is $\psi_{k}(x_{0},x_{k}) : = \frac{x_{0}x_{k}}{k^{\alpha}}$. It is known that, for $\alpha >2$ the 
Bowen-Walters' condition holds and thus there is a unique equilibrium state~\cite{Dyson2}. 

We will assume the following condition.

\begin{cond}\label{Condi}
Let $p:\rr\to\rr$ be a function such that $\frac{p(r)}{r}\to\infty$ as $r\to\infty$, we consider
pair-wise interactions $\psi_{k}$ satisfying that
\begin{equation*}
\sum_{r=0}^{\infty}p(r)\lVert \psi_{r}\rVert <\infty.
\end{equation*}
\end{cond}

\begin{rem}
Observe that this condition is slightly stronger than the Bowen-Walters' condition and the Ruelle's condition~\cite{Ruelle}.
We will emphasize the advantage of considering this condition instead.
\end{rem}

\subsection{Gibbs measures and equilibrium states}
The probability measure $\mu\in \mathcal{M}_{T}(X)$ is said to be an equilibrium state for $\phi$ if 
\[
h(\mu) + \int \phi\dd\mu = \sup_{\nu \in\mathcal{M}_{T}(X)}\{h(\nu) + \int \phi\dd\nu \}.
\]
Such a definition is indeed the well-known \emph{variational principle}. As mentioned above, 
the variational principle was borrow from statistical mechanics to generalize the concept of
equilibrium state as the measure that ``maximizes'' the Helmholtz free energy. In the context of
the thermodynamical formalism, the ``maximum'' of the free energy is given the name of 
\emph{topological pressure} (see definition below). For lattice systems it is known that the 
measures fulfilling the variational principle are the Gibbs measures~\cite{LR}. 

\begin{defi}[Gibbs Measures]
Let $(X,T)$ be the full-shift and $\phi:X\to\rr$ be a continuous potential. 
The measure $\mu_{\phi}$ associated to the potential $\phi$ is a Gibbs measure 
if there exists constants $C>1$ and $P(\phi)>0$ such that for all $x\in X$
\[
\frac{\mu_{\phi}[x_{0}^{n-1}]}{\exp(S_{n}\phi(y) + nP(\phi))} = C^{\pm},
\] 
for all $y\in[x_{0}^{n-1}]$. Where $S_{n}\phi(y) :=\sum_{k=0}^{n-1}\phi(T^{k}y)$. The constant $P(\phi)$ is called the \emph{topological pressure} and 
it is such that
\[
P(\phi) = h(\mu) + \int \phi\dd\mu
\]
where $\mu$ is the measure that fulfills the variational principle.
\end{defi}

\subsection{Markov approximations to Gibbs measures} In this work we use the same 
technique as the one implemented in~ \cite{ChRU} to characterize the Gibbs 
measures associated to infinite range potentials. The strategy goes as follows. 
We will use the fact that every Gibbs measure can be obtained as the limit  
of a sequence of periodic measures, defined as
\begin{defi}[Periodic Measures]
Let $p\in\nn$ be a positive integer and denote by $\Per_{p}(X)$ the set of all periodic sequences $x\in X$ of period $p$. 
For every $n\in\nn$ and $p>n $, the periodic measure of period $p$ is defined by 
\begin{equation}\label{periodic}
\P^{(p)}[a_{0}^{n-1}]:= \frac{\sum_{y\in\Per_{p}(X)\cap[a_{0}^{n-1}]}
e^{S_{n}\phi(y)}}{\sum_{y\in\Per_{p}(X)}e^{S_{n}\phi(y)}}.
\end{equation}
\end{defi}

In order to explicitly build up the sequence of measures having as a limit the Gibbs 
measure for infinite range potential $\phi : X \to \mathbb{R}$ it is necessary an intermediate step. 
We need to construct a Gibbs measure for potentials of range $r$. The latter will be taken as ``approximations'' to the infinite range potential we are interested in.  The $r$-range approximation $\phi_{r} : X \to \mathbb{R}$ to the potential $\phi$ is defined as,
\begin{equation}
\phi_{r}(x) := \sum_{k=0}^{r-1}\psi_{k}(x_{0},x_{k}).
\label{r-poten}
\end{equation}
\noindent
In this case, the potential $\phi_r$ admits a unique equilibrium measure on the full shift $X$, 
which is also a ($r$-step) Markov measure. This measure can be characterized by means of the concept of \emph{transfer matrix}. 
Given the potential $\phi_{r}$, we will denote by $\tilde{\phi}_{r}$ 
a function from $A^{r}$ to $\mathbb{R}$ as $ \tilde{\phi}_{r} (x_0,x_1, \dots, x_{r-1}) = \phi_{r} (x) $ for every $x = x_0 x_1x_2\dots \in X$. 
Then we define the transition matrix $M_{r}:A^{r}\times A^{r}\to \rr^{+}$ associated to $\phi_r$, by
\begin{equation}\label{transmatrix}
M_{{r}}(a,b):=
\begin{cases}
\exp{\phi_{r}(ab_{r-1})} &\quad \mbox{if}\ a_{1}^{r-1}=b_{0}^{r-2},\\
0 &\quad \mbox{otherwise}.
\end{cases}
\end{equation}
In the last expression $ab_{r-1}$ means the concatenation of the word $a$ and the last letter
of the word $b$. We will denote by $L_{{r}}: A \to \mathbb{R}$ and $R_{{r}}: A \to \mathbb{R}$ 
the corresponding left and right eigenvectors of $M_r$. Then the measure defined by 
\[
\mu_{\phi_{r}}[a_{0}^{n-1}]:= L_{{r}}(a_{0}^{r-1})
\frac{\prod_{j=0}^{n-r}M_{{r}}(a_{j}^{j+r-1},a_{j+1}^{j+r})}{\rho_{{r}}^{n-r+1}}
R_{{r}}(a_{n-r+1}^{n-1})
\]
for each $x\in X$ and $n\in\nn$ such that $n\geq r$, is a shift-invariant
probability measure. Moreover, $\mu_{\phi_{r}}$  is the unique Gibbs measure associated to the  potential $\phi_{r}$~\cite{ChU}.

\section{Results}
Our main result is the following.

\begin{thm}\label{Main}
Consider the infinite-range interaction  $\phi$ given by \eqref{poten} and 
let $\phi_{r}$ be its finite-range approximation.  
Let $\mu_{\phi_{r}}$ be the Markov measure associated to $\phi_{r}$. 
If $\phi$ satisfies the condition \ref{Condi}, 
then $\mu_{\phi_{r}}$ converges to $\mu_{\phi}$ in the projective sense. 
As a consequence $\mu_{\phi_{r}}$ converges to $\mu_{\phi}$ in the weak-$*$ topology.
\end{thm}

The technique used in this paper plus the condition assumed on the potential allows us to obtain a stretched exponential rate of mixing for the limiting measure. In our context this result enhances the rate of mixing obtained in~\cite{ChRU} which is of order $\sqrt{s}$. Our result is the following.

\begin{thm}\label{Mixing}
Assume that the infinite-range potential $\phi$ satisfies the condition~\ref{Condi}, then there exist constants $C>1$, 
$c>0$ and $\xi\in(0,1)$ such that for any $a, b \in A^{r}$ there exists $s^{*} = s(a,b)$ for which
\[
\left\lvert \frac{\mu_{\phi}([a]\cap T^{-s}[b])}{\mu_{\phi}[a]\mu_{\phi}[b]} -1 \right\rvert\leq C s e^{- c s^{\xi}},
\]
for all $s\geq s^{*}$.
\end{thm}

Finally we also are able to obtain a result for the convergence of the entropies of
the approximating Markov measures to the entropy of the equilibrium state.

\begin{thm}\label{Entropy}
Let $\phi: X \to \mathbb{R}$ be a long-range potential as defined above fulfilling  condition~\ref{Condi}. 
Then, there are constants $C>0$ and $\epsilon>0$ such that
\[
|h(\mu_\phi) - h(\mu_{\phi_r})| \leq C r^{-\epsilon}.
\]
\end{thm}

\subsection{Concluding Remarks}

Our main contribution to the study of the Gibbs measures on one-dimensional lattices  with long range pair interactions (with a finite state space) is the improvement of previous results within the framework of thermodynamical formalism. Particularly we proved projective convergence of a sequence of finite range Markov approximations to the unique Gibbs measure (Theorem~\ref{Main}) corresponding to a given potential. We also proved that the latter is mixing with a stretched exponential rate (Theorem~\ref{Mixing}). Finally we also proved that the entropies of the finite range approximations also converges, a result which is a direct consequence of the projective convergence (Theorem~\eqref{Entropy}). 
For the particular case of polynomially decaying pair interactions, i.e. $\psi_k = -\beta J k^{-\alpha}$ (with $\beta $ as the inverse temperature and $J$ as a ``coupling" constant) it is known that for $\alpha>2$  there is a unique ergodic measure for all the values of $\beta$~\cite{dyson}. Since this potential comply with Condition~\ref{Condi} for $\alpha>2$, our analysis let us conclude that for all $\beta \geq 0$ the Markov approximations converge in the projective sense. Moreover, the above theorems lets us obtain explicit formulae for the mixing rate and for the convergence rate of the entropy. 
On the other hand for the case $\alpha \leq 2$ it is known that for some $\beta_c$ all the values of $\beta < \beta_c$ the Gibbs measure $\mu_{\beta \phi}$ is the convex combination of two ergodic measures. In such a case the weak-$*$ limit does exist and has two ergodic components. However, the Markov approximations cannot converge projectively by a theorem about the projective limit  of ergodic measures  due to~\cite{liliana}.
Indeed, according to~\cite{liliana} the loss of projective convergence may be a signature of non-uniqueness of ergodic measures. They prove that if a sequence of ergodic measures converge in the projective sense then the limit is ergodic. In our work we proved that for the case of one-dimensional lattices with, to some extent, arbitrary pair interactions the Markov approximations converges projectively to the unique Gibbs measure when the interaction fulfills Condition~\ref{Condi}. If Condition~\ref{Condi} does not hold we have the possibility that the Gibbs measure might be decomposed in two or more ergodic measures. As mentioned above, in this case such a limit measure cannot be reached by a sequence of Markov approximations because they are ergodic (pure states) and the ``limit measure'' it is not. 
It would be interesting to see if for the case $\beta>\beta_c$ and $1< \alpha \leq 2$ (for which the Gibbs measure is still uniquely ergodic~\cite{FS}) we can prove projective convergence of Markov approximations, a fact that could give us a criterium to detect phase transitions.


\section{Proofs}

In order to give the proof of theorem \ref{Main}, let us first introduce the following definition.
Let $n,r\in \nn$ be fixed and $n\geq r$. 
Consider $a_{0}^{n-1}\in A^{n}$. For every $p>n+r$ the
$r$-Markov approximation $\P_{r}^{(p)}$ of the periodic measure 
\eqref{periodic} is given by
\[
\P^{(p)}_{r}[a_{0}^{n-1}] := \frac{\sum_{y\in\Per_{p}(X)\cap[a_{0}^{n-1}]}
e^{S_{n}\phi_{r}(y)}}{\sum_{y\in\Per_{p}(X)}e^{S_{n}\phi_{r}(y)}},
\]
where $\phi_{r}$ is the $r$-range approximation of $\phi$, and it is given by \eqref{r-poten}.
This measures are also called elementary Gibbs measures. We have the following lemma.

\begin{lemma}\label{PeriodApp}
Let $\mu_{\phi_{r}}$ be the $r$-range Markov approximation of the measure $\mu_{\phi}$.
And let $\P_{r}^{(p)}$ be the $r$-range approximation of the periodic measure $\P^{(p)}$. 
Then there exist constants $C_{r}>0$ and $\eta<1$ for all $a_{0}^{n-1}\in A^{n}$, such that
\begin{equation*}
\mu_{\phi_{r}}[a_{0}^{n-1}]
=\P_{r}^{(p)}[a_{0}^{n-1}]\cdot
\exp\left(\pm 12 r C_{r} e^{2 C_{r}}\cdot
\eta^{\frac{p-n}{r}-2}\right).
\end{equation*}
\end{lemma}

\noindent
Let us assume for the moment this lemma, the proof will be given after the proof of the theorem~\ref{Main}.
The explicit expressions for the constants $C_{r}$ and $\eta$ are given in the proof.

\subsection{Proof of Theorem \ref{Main}}

Consider two consecutive Markov approximations. 
We prove that they accumulate towards a limit measure as the range of the approximation 
diverges. The strategy is to use the periodic approximation of the Markov measure as follows.
Using lemma \ref{PeriodApp} for $(r+1)$-range interactions, one obtains that
\[
\mu_{\phi_{r+1}}[a_{0}^{n-1}]= \P_{r+1}^{(p)}[a_{0}^{n-1}]\cdot
\exp\left(\pm (r+1)C_{r+1}e^{2 C_{r+1}}\cdot \eta^{\frac{p-q}{r+1} -2}\right),
\]
where $\eta :=1 - e^{-2\sum_{k=1}^{\infty}k\lVert\psi_{k}\rVert}$ and 
$q:= \max\{n,r+1\}$. For the sake of clarity, let us drop the indices on $a_{0}^{n-1}$. 
We compare two consecutive Markov measures,
\begin{align*}
\frac{\mu_{\phi_{r}}[a]}{\mu_{\phi_{r+1}}[a]} = &
\frac{\P_{r}^{(p)}[a]}{\P_{r+1}^{(p)}[a]}\times
\frac{\exp\left(\pm 12 r C_{r}e^{2 C_{r}}\cdot\eta^{\frac{p-n}{r} -2}\right)}
{\exp\left(\pm 12(r+1)C_{r+1}e^{2 C_{r+1}}\cdot\eta^{\frac{p-q}{r+1} -2}\right)}\\
=&
\frac{\P_{r}^{(p)}[a]}{\P_{r+1}^{(p)}[a]}\times
\exp\left(\pm24(r+1)C_{r+1}e^{2 C_{r+1}}\cdot\eta^{\frac{p-q}{r+1}-2} \right).
\end{align*}

\noindent
In order to have a bound depending only on the interaction we compare 
two consecutive periodic measures. Since
$
\lvert S_{p}\phi_{r}(y) - S_{p}\phi_{r+1}(y)\rvert = 
\lvert S_{p}\psi_{r+1}(y_{0},y_{r+1})\rvert \leq (p+1)\lVert \psi_{r+1}\rVert
$
we obtain that
\[
\frac{\P_{r}^{(p)}[a]}{\P_{r+1}^{(p)}[a]} = 
e^{\pm2(p+1)\lVert\psi_{r+1}\rVert}.
\]
Finally, putting together these previous bounds yields
\begin{align*}
\frac{\mu_{\phi_{r}}[a]}{\mu_{\phi_{r+1}}[a]} = &
\exp\left(\pm2(p+1)\lVert\psi_{r+1}\rVert\right)\cdot
\exp\left(\pm24(r+1)C_{r+1}e^{2 C_{r+1}}\eta^{\frac{p-q}{r+1}-2} \right)\\
=& \exp\left(\pm\big(2(p+1)\lVert\psi_{r+1}\rVert+ 
24(r+1)C_{\infty}e^{2 C_{\infty}}\eta^{\frac{p-n-r-1}{r+1}-2}\big) \right),
\end{align*}
where $C_{\infty} = \sum_{k=1}^{\infty}k\lVert\psi_{k}\rVert$. 
Let $p:=p(s)$, then for any word $a\in A^{n}$ and any 
$r'>r\in\nn$, one has that
\[
\frac{\mu_{\phi_{r}}[a]}{\mu_{\phi_{r'}}[a]} = 
\exp\Big(\pm \Big(2\sum_{s= r}^{r'}p(s)\lVert\psi_{s}\rVert + 
24R\sum_{s= r}^{r'}s\eta^{\frac{p(s)}{s}} \Big) \Big),
\]
where $R:= C_{\infty}e^{2 C_{\infty}}$. Since $\eta<1$ and assuming
the condition \ref{Condi} the right hand side of the previous expression
tends to zero as $r$ diverges. Thus, the limit
$\mu[a] :=\lim_{r\to\infty}\mu_{\phi_{r}}[a]$, exists for 
any word $a\in A^{n}$, proving the weak-$*$ convergence. Moreover, one has that
\begin{equation}\label{rmarkov-lim}
\frac{\mu_{\phi_{r}}[a]}{\mu[a]} = 
\exp\Big(\pm \Big(2\sum_{s\geq r}p(s)\lVert\psi_{s}\rVert + 
24R\sum_{s\geq r}s\eta^{\frac{p(s)}{s}} \Big) \Big),
\end{equation}
for every $r\in\nn$ and any word $a\in A^{n}$. So one also has that $\mu_{\phi_{r}}[a]/\mu_{\phi}[a] = e^{\pm n\varepsilon_{r}}$, where $n$ is the length of the word $a$, and $\varepsilon_{r}$ is giving by
\[
\varepsilon_{r}:=  2\sum_{s\geq r}p(s)\lVert\psi_{s}\rVert + 
24R\sum_{s\geq r}s\eta^{\frac{p(s)}{s}} .
\]
Which proves the projective convergence of the Markov approximations $\mu_{\phi_{r}}$ to $\mu$.
The limiting measure $\mu$ turns to be absolutely continuous to $\mu_\phi$ since 
we use the elementary Gibbs measures to approximate the Markov ones. 
The proof of this statement is found in \cite[Section 6.3.3.]{ChU}.
And that finishes the proof.
\qed

\subsubsection{Proof of lemma \ref{PeriodApp}.}

Here we use that every Markov measure can be seen as a limit of measures 
with support on the periodic points as it was proved in~\cite[Appendix 6.3.1.]{ChU}. 
From the calculations in that paper, one has explicit bounds for the Markov 
measure $\mu_{r}$ in terms of the finite-range periodic approximation $\P_{r}^{(p)}$ as follows,

\begin{equation*}
\mu_{\phi_{r}}[a] = \P_{r}^{(p)}[a]
\exp\left(\pm\frac{rD_{0}}{1-\eta}\eta^{\frac{p-n}{r}-2} \right),
\end{equation*}
where $\eta:=\eta(M_{r}^{r})$, is the Birkhoff's coefficient of the matrix $M_{r}^{r}$ 
and $D_{0}$ is a constant bounded by
\begin{align*}
D_{0}\leq& 2\max_{b,c,c'\in A^{r}}\log\left(\frac{M_{r}^{r}(c,b)M_{r}^{r+1}(c',b)}
{M_{r}^{r+1}(c,b)M_{r}^{r}(c',b)} \right)\\
=& \frac{e^{ S_{r}\phi_{r}(cb)}e^{ S_{r+1}\phi_{r}(c'b)}}
{e^{ S_{r+1}\phi_{r}(cb)}e^{ S_{r}\phi_{r}(c'b)}}.
\end{align*}
Using that
\[
\lvert S_{r}\phi_{r}(bc') - S_{r+1}\phi_{r}(bc')\rvert\leq 
3\sum_{k=1}^{r}k\lVert \psi_{k}\rVert,
\]
for $b\in A^{r}$ and $c'\in A$. One gets
\[
D_{0}\leq 2 C_{r},
\]
where $C_{r}:=\sum_{k=1}^{r}k\lVert\psi_{k}\rVert$. And so, one has
\[
\mu_{\phi_{r}}[a]=
\P_{r}^{(p)}[a]\cdot
\exp\left(\pm 12 r C_{r} e^{2 C_{r}}\times
\eta^{\frac{p-n}{r}-2}\right),
\]
which proves the lemma for $\eta =1 - e^{-2\sum_{k=1}^{\infty}k\lVert\psi_{k}\rVert} $ and $C_{r} = \sum_{k=1}^{r}k\lVert \psi_{k}\rVert$.
\qed

\subsection{Proof of theorem \ref{Mixing}. }
Here we mimic the proof of theorem 3.3 in \cite{ChRU}.
As before we make use of the periodic approximation of the limiting measure through the
following lemma.

\begin{lemma}\label{Lemitita}
Let $\P_{r}^{(p)}$ be the $r$-range approximation of the periodic measure $\P^{(p)}$. 
Let $s, r\in \nn$. For any $a, b\in A^{r}$ one has that
\begin{equation}\label{lemitita}
\P_{r}^{(p)}([a]\cap T^{-s}[b]) = \P_{r}^{(p)}[a]\P_{r}^{(p)}[b] 
\times \exp\left(\pm 4\frac{rD}{1-\eta}\eta^{\lfloor\frac{s}{r}\rfloor} \right),
\end{equation}
for a constant $D>0$ and where $\eta$ is the same constant as in lemma \ref{PeriodApp}.
\end{lemma}
\noindent
From inequality \eqref{rmarkov-lim} applied to the limiting measure one can easily obtain that
\begin{align*}
\lvert \mu_{\phi}([a]\cap T^{-s}[b])-\mu_{\phi}[a]\mu_{\phi}[b]\rvert \leq &
\lvert \P_{r}^{(p)}([a]\cap T^{-s}[b]) - \P_{r}^{(p)}[a]\P_{r}^{(p)}[b]\rvert \times\\
&\exp\Big( 2 \Big[ 2\sum_{k\geq r}p(k)\lVert \psi_{k}\rVert + 
24R\sum_{k\geq r}k\eta^{\frac{p(k)}{k}} \Big] \Big).
\end{align*}
Using the inequality \eqref{lemitita} from our previous lemma, yields
\begin{align*}
\left\lvert \frac{\mu_{\phi}([a]\cap T^{-s}[b])}{\mu_{\phi}[a]\mu_{\phi}[b]} - 1\right\rvert \leq &
\left\lvert \exp\Big( 4 r\frac{D}{1-\eta}\eta^{\lfloor \frac{s}{r}\rfloor} \Big) -1 \right\rvert \times\\
&\exp\Big( 2 \Big[ 2\sum_{k\geq r}p(k)\lVert \psi_{k}\rVert + 
24R\sum_{k\geq r}k\eta^{\frac{p(k)}{k}} \Big] \Big).
\end{align*}

\noindent
By the hypothesis made on the potential, the second factor in the 
right-hand side of the last inequality can be bounded by above by a 
constant $C>1$ for every $r\geq1$.
The larger $r$ we take, the smaller the sufficient upper bound $C$ becomes.

Next, let us consider $r=r(s)$ such that $r(s) = o(s)$. Then there
exists a constant $0<\xi<1$ such that for every pair $a,b\in A^{r}$ there exists a $s^{*}\in\nn$ for which
\begin{align*}
\left\lvert \frac{\mu_{\phi}([a]\cap T^{-s}[b])}{\mu_{\phi}[a]\mu_{\phi}[b]} - 1\right\rvert \leq Cs\eta^{s^{\xi}},
\end{align*}
for every $s\geq s^{*}$. 
Now, by introducing an adequate positive constant $c$ we change the base and it is possible to rewrite the previous inequality into the form 
\begin{align*}
\left\lvert \frac{\mu_{\phi}([a]\cap T^{-s}[b])}{\mu_{\phi}[a]\mu_{\phi}[b]} - 1\right\rvert \leq Cs e^{-cs^{\xi}},
\end{align*}
again for every $s\geq s^{*}$. 
And that finishes the proof of the theorem.
\qed

\subsubsection{Proof of lemma \ref{Lemitita}.}
First, let $r\in\mathbb{N}$, $p>2r$ and let $M_{r}$ be the transition matrix given by \eqref{transmatrix}. 
Consider any $a\in A^{r}$, we have
\[
\P_{r}^{(p)}[a] = \frac{\sum_{x\in\Per_{p}(X)\cap[a]}e^{S_{p}\phi_{r}(x)}}{\sum_{x\in \Per_{p}(X)}e^{S_{p}\phi_{r}(x)}} 
= \frac{e_{a}^{\dagger}M_{r}^{p}e_{a}}{\sum_{x\in A^{r}} e_{x}^{\dagger}M_{r}^{p}e_{x}},
\]
where $e_{x}$ is a vector with an entry equal to 1 at position $x$ and 0's elsewhere.
\noindent
For any $s\in\nn$, it is easy to see that
\[
\P_{r}^{(p)}([a]\cap T^{-s}[b])= \sum_{w\in A^{s}}\P_{r}^{(p)}[awb].
\]

\noindent
Next, chose an arbitrary but fixed $w\in A^{s}$. We are able to write $\P_{r}^{(p)}[awb]$ in terms of the transition matrix $M_{r}$, as follows,
\begin{equation*}
\P_{r}^{(p)}[awb]=
\frac{e^{\dagger}_{a}M_{r}^{p-a}e_{wb}e^{\dagger}_{wb}M_{r}^{s}e_{a}}{\sum_{a}\sum_{wb}e^{\dagger}_{a}M_{r}^{p-s}e_{wb}e_{wb}^{\dagger}M_{r}^{s}e_{a}}.
\end{equation*}
Here we use the enhanced Perron-Frobenius theorem \cite[Corollary 2.16.]{ChU}, yielding
\begin{equation*}
\begin{split}
\P_{r}^{(p)}[awb] =& \frac{e^{\dagger}_{a}(L^{\dagger}e_{wb})
\rho^{p-s}R\cdot e^{\dagger}_{wb}(L^{\dagger}e_{a})\rho^{s}R}
{\sum_{a}\sum_{wb} e^{\dagger}_{a}(L^{\dagger}e_{wb}) 
\rho^{p-s}R\cdot e^{\dagger}_{wb}(L^{\dagger}e_{a})\rho^{s}R} \times \\
&\exp\left(\pm2\frac{r\delta(e_{wb},Fe_{wb})}{1-\eta}\eta^{\lfloor \frac{p-s}{r}\rfloor}\right)\cdot
\exp\left(\pm2\frac{r\delta(e_{a},Fe_{a})}{1-\eta}\eta^{\lfloor\frac{s}{r}\rfloor} \right).
\end{split}
\end{equation*}
Where $F$ is the contraction defined on the simplex and $\delta$ is a distance defined on the simplex, both are explicitly defined in~\cite[Appendix 1.]{ChRU} and~\cite[Appendix 6.1.]{ChU}.
\noindent
Since $L^{\dagger}e_{x} = L(x)$ and $e^{\dagger}_{x}R = R(x)$ one has that
\begin{align*}
\P_{r}^{(p)}[awb]=& 
\frac{L(wb)R(a)\cdot L(a)R(wb)}{\sum_{a}\sum_{wb} L(wb)R(a)\cdot L(a)R(wb)}\times \\
& \exp\left(\pm2\frac{r\delta(e_{wb},Fe_{wb})}{1-\eta}\eta^{\lfloor \frac{p-s}{r}\rfloor}\right)\cdot
\exp\left(\pm2\frac{r\delta(e_{a},Fe_{a})}{1-\eta}\eta^{\lfloor\frac{s}{r}\rfloor} \right)\\
= & \P_{r}^{(p)}(a)\P_{r}^{(p)}(wb) \times \\
&\exp\left(\pm2\frac{r\delta(e_{wb},Fe_{wb})}{1-\eta}\eta^{\lfloor \frac{p-s}{r}\rfloor}\right)\cdot
\exp\left(\pm2\frac{r\delta(e_{a},Fe_{a})}{1-\eta}\eta^{\lfloor\frac{s}{r}\rfloor} \right).
\end{align*}
Where in the last inequality we used the fact that $\P$ is a probability measure. Finaly one obtains that
\begin{align*}
\P_{r}^{(p)}([a]\cap T^{-s}[b]) = &\sum_{w\in A^{s}}\P_{r}^{(p)}(a)\P_{r}^{(p)}(wb)\times \\
 &\exp\left(\pm2\frac{r\delta(e_{wb},Fe_{wb})}{1-\eta}\eta^{\lfloor \frac{p-s}{r}\rfloor}\right)\cdot
\exp\left(\pm2\frac{r\delta(e_{a},Fe_{a})}{1-\eta}\eta^{\lfloor\frac{s}{r}\rfloor} \right)\\
= & \P_{r}^{(p)}(a)\P_{r}^{(p)}(b) \times \exp\left(\pm 4\frac{rD}{1-\eta}\eta^{\lfloor\frac{s}{r}\rfloor} \right),
\end{align*}
for some constant $D>0$. We remind that in our particular case the Birkhoff's coefficient 
$\eta$ is giving by $\eta = 1- e^{-2\sum_{k=1}^{\infty}k\lVert \psi_{k}\rVert}$.
\qed

\subsection{Proof of theorem \ref{Entropy}.}

First notice that given $r>0$ and given $a\in \mathcal{A}^{k}$  with $k>0$ we have that for every $x,y\in [a]$
\[
|\phi(x)-\phi_r(y)| = \sum_{n=r}^\infty \psi_{n}(x_0,x_n) 
\]
then we have that, for every $k>0$ the sums difference $S_{k}\phi(x)-S_{k}\phi_r(y)$ can be bounded by
\begin{eqnarray}
|S_k\phi (x) - S_{k}\phi_r(y)| &=& \sum_{n=r}^\infty \sum_{j=0}^{k-1} \psi_{n}(x_j, x_{j+n}) 
\nonumber  
\\
                          &\leq& k\sum_{n=r}^\infty  || \psi_{n} ||. \nonumber 
\end{eqnarray}
Since $\psi$ has summable variation it is clear that the last sum is finite and goes to zero as $r\to \infty$. Moreover, for polynomially decaying potential, such can be bounded by 
\[
\sum_{n=r}^\infty  || \psi_{n} || \leq C^\prime r^{-\epsilon }
\]
for some $\epsilon>0$. The last constant depends on  how the  pair potential $\psi_k$ decays with the ``distance'' $k$.  
From the above we can easily see that
\[
\exp\left\{ S_k\phi(a^*)\right\}  = \exp\left\{ S_k\phi_r(a^*) \right\} \exp\left( \pm k C^\prime r^{-\epsilon}  \right).
\]
This result lets us bound the difference of topological pressures as follows
\begin{eqnarray}
0 \leq |P(\phi,X) -  P(\phi_r,X)| &\leq& \frac{1}{k} \log \bigg( \frac{\sum_{a\in \mathrm{Per}_{k}}  \exp\left\{ S_k\phi(a^*)\right\}    }{\sum_{b\in \mathrm{Per}_{k}}  \exp\left\{ S_k\phi_r(b^*)\right\}   } \bigg)
\nonumber 
\\
&\leq& C^\prime r^{-\epsilon} 
\nonumber
\end{eqnarray}
where $a^*$ ($b^*$ respectively) is some point $X$ belonging to $[a]$ ($[b]$ respectively). 

Now, since both $\mu_{\phi_r}$ and $\mu_{\phi}$ satisfy  the variational principle~\cite{Bow} we have that 
\[
P(\phi,X) = \int_X \phi d\mu_\phi + h(\mu_\phi) \quad \mbox{and} \quad P(\phi_r,X) = \int_X \phi_r d\mu_{\phi_r} + h(\mu_{\phi_r}),
\]
which lets us write
\begin{equation}
| h(\mu_\phi) - h(\mu_{\phi_r})| \leq |P(\phi,X) -  P(\phi_r,X)|  + \bigg| \int_X \phi d\mu_\phi - \int_X \phi_r d\mu_{\phi_r} \bigg|.
\label{eq:upperbound_dh}
\end{equation}
Now, by~\eqref{rmarkov-lim} we know that
\[ 
\mu_\phi = \mu_{\phi_r} \exp(\pm \varepsilon_{r}),
\]
which lets us write
\[
\bigg| \int_X \phi d\mu_\phi - \int_X \phi_r d\mu_{\phi_r} \bigg| \leq \left( 1- \exp( - \varepsilon_{r} ) \right) || \phi ||.
\]
It is clear that the right-hand side of the above equation vanishes exponentially, which lets us conclude that the left-hand side of inequality~\eqref{eq:upperbound_dh} is bounded by $C r^{-\epsilon}$ choosing an appropriate constant $C >0$,
\[
| h(\mu_\phi) - h(\mu_{\phi_r})| \leq C r^{-\epsilon},
\]
which proves the theorem.
\qed

\bibliographystyle{plain}
\bibliography{biblio}

\def\cprime{$'$} \def\cprime{$'$}
\begin{thebibliography}{10}

\bibitem{Bow2}
Rufus Bowen.
\newblock Some systems with unique equilibrium states.
\newblock {\em Math. Systems Theory}, 8(3):193--202, 1974/75.

\bibitem{Bow}
Rufus Bowen.
\newblock {\em Equilibrium states and the ergodic theory of {A}nosov
  diffeomorphisms}, volume 470 of {\em Lecture Notes in Mathematics}.
\newblock Springer-Verlag, Berlin, revised edition, 2008.
\newblock With a preface by David Ruelle, Edited by Jean-Ren{\'e} Chazottes.

\bibitem{ChRU}
J.-R. Chazottes, L.~Ramirez, and E.~Ugalde.
\newblock Finite type approximations of {G}ibbs measures on sofic subshifts.
\newblock {\em Nonlinearity}, 18(1):445--463, 2005.

\bibitem{ChU}
J.-R. Chazottes and E.~Ugalde.
\newblock On the preservation of {G}ibbsianness under symbol amalgamation.
\newblock In {\em Entropy of hidden {M}arkov processes and connections to
  dynamical systems}, volume 385 of {\em London Math. Soc. Lecture Note Ser.},
  pages 72--97. Cambridge Univ. Press, Cambridge, 2011.

\bibitem{dobrushin1968gibbsian}
R.~L. Dobru{\v{s}}in.
\newblock Gibbsian random fields for lattice systems with pairwise
  interactions.
\newblock {\em Funkcional. Anal. i Prilo\v zen.}, 2(4):31--43, 1968.

\bibitem{dobrushin1968problem}
R.~L. Dobru{\v{s}}in.
\newblock The problem of uniqueness of a {G}ibbsian random field and the
  problem of phase transitions.
\newblock {\em Funkcional. Anal. i Prilo\v zen.}, 2(4):44--57, 1968.

\bibitem{dyson}
Freeman~J. Dyson.
\newblock Existence of a phase-transition in a one-dimensional {I}sing
  ferromagnet.
\newblock {\em Comm. Math. Phys.}, 12(2):91--107, 1969.

\bibitem{Dyson2}
Freeman~J. Dyson.
\newblock Non-existence of spontaneous magnetization in a one-dimensional ising
  ferromagnet.
\newblock {\em Communications in Mathematical Physics}, 12:212--215, 1969.

\bibitem{FS}
J.~Fr{\"o}hlich and T.~Spencer.
\newblock The phase transition in the one-dimensional {I}sing model with
  {$1/r^{2}$} interaction energy.
\newblock {\em Comm. Math. Phys.}, 84(1):87--101, 1982.

\bibitem{LR}
O.~E. Lanford, III and D.~Ruelle.
\newblock Observables at infinity and states with short range correlations in
  statistical mechanics.
\newblock {\em Comm. Math. Phys.}, 13:194--215, 1969.

\bibitem{Ruelle}
D.~Ruelle.
\newblock Statistical mechanics of a one-dimensional lattice gas.
\newblock {\em Comm. Math. Phys.}, 9:267--278, 1968.

\bibitem{Sten}
{\"O}rjan Stenflo.
\newblock Uniqueness in {$g$}-measures.
\newblock {\em Nonlinearity}, 16(2):403--410, 2003.

\bibitem{liliana}
L.~Trejo and E.~Ugalde.
\newblock In preparation.

\bibitem{Walters}
Peter Walters.
\newblock Ruelle's operator theorem and {$g$}-measures.
\newblock {\em Trans. Amer. Math. Soc.}, 214:375--387, 1975.

\end{thebibliography}

\end{document}